\documentclass{emulateapj}

\shorttitle{Ultra-Faint dSph Metallicities}
\shortauthors{Kirby et al.}
\slugcomment{Accepted for publication in ApJL}

\begin{document}
\newcommand{\deimos}{\textsc{deimos}}
\newcommand{\teff}{$T_{\rm{eff}}$}
\newcommand{\mathteff}{T_{\rm{eff}}}
\newcommand{\logg}{$\log g$}
\newcommand{\feh}{[Fe/H]}
\newcommand{\mathfeh}{\rm{[Fe/H]}}
\newcommand{\afe}{[$\alpha$/Fe]}
\newcommand{\mathafe}{\rm{[\alpha/Fe]}}

\title{Uncovering Extremely Metal-Poor Stars in the Milky Way's
  Ultra-Faint Dwarf Spheroidal Satellite Galaxies\altaffilmark{1}}

\author{Evan~N.~Kirby\altaffilmark{2},
  Joshua~D.~Simon\altaffilmark{3}, Marla~Geha\altaffilmark{4}, Puragra
  Guhathakurta\altaffilmark{2}, and Anna Frebel\altaffilmark{5}}

\altaffiltext{1}{Data herein were obtained at the W.~M. Keck
  Observatory, which is operated as a scientific partnership among the
  California Institute of Technology, the University of California,
  and NASA.  The Observatory was made possible by the generous
  financial support of the W.~M. Keck Foundation.}
\altaffiltext{2}{University of California Observatories/Lick
  Observatory, Department of Astronomy \& Astrophysics, University of
  California, Santa Cruz, CA 95064; \texttt{ekirby@ucolick.org}}
\altaffiltext{3}{Department of Astronomy, California Institute of
  Technology, 1200 E. California Blvd., MS 105-24, Pasadena, CA 91125}
\altaffiltext{4}{Astronomy Department, Yale University, P.O. Box
  208101, New Haven, CT 06520}
\altaffiltext{5}{McDonald Observatory, University of Texas, Austin, TX
  78712}

\keywords{galaxies: dwarf --- galaxies: abundances --- Galaxy: halo
  --- Galaxy: evolution --- Local Group}

%%%%%%%%%%%%%%%%%%%%%%%%%%%%%%%%%
%%%%%%%%%    ABSTRACT    %%%%%%%%
%%%%%%%%%%%%%%%%%%%%%%%%%%%%%%%%%

\begin{abstract}

We present new metallicity measurements for 298 individual red giant
branch stars in eight of the least luminous dwarf spheroidal galaxies
(dSphs) in the Milky Way (MW) system.  Our technique is based on
medium resolution Keck/DEIMOS spectroscopy coupled with spectral
synthesis.  We present the first spectroscopic metallicities at
$\mathfeh < -3.0$ of stars in a dwarf galaxy, with individual stellar
metallicities as low as $\mathfeh = -3.3$.  Because our
\feh\ measurements are not tied to empirical metallicity calibrators
and are sensitive to arbitrarily low metallicities, we are able to
probe this extremely metal-poor regime accurately.  The metallicity
distribution of stars in these dSphs is similar to the MW halo at the
metal-poor end.  We also demonstrate that the luminosity-metallicity
relation previously seen in more luminous dSph galaxies ($M_V = -13.4$
to $-8.8$) extends smoothly down to an absolute magnitude of $M_V =
-3.7$.  The discovery of extremely metal-poor stars in dSphs lends
support to the $\Lambda$CDM galaxy assembly paradigm wherein dwarf
galaxies dissolve to form the stellar halo of the MW.

\end{abstract}

%%%%%%%%%%%%%%%%%%%%%%%%%%%%%%%%%
%%%%%%%%%   SECTION 1   %%%%%%%%%
%%%%%%%%%%%%%%%%%%%%%%%%%%%%%%%%%

\section{Introduction}
\label{sec:intro}
In the hierarchical theory of the assembly of galaxy halos
\citep{whi78,sea78}, dwarf galaxies interact gravitationally with
their hosts, shedding stars, losing gas, and eventually tidally
dissolving into the diffuse halo.  Recent numerical simulations and
semi-analytic models \citep[e.g.,][]{bul05} suggest that a Milky
Way-like halo can be explained entirely by disrupted satellites in the
$\Lambda$CDM paradigm.  
%Numerical simulations show that disrupted satellites leave an
%observable kinematic imprint on a galactic halo \citep{hel99}.  Other
%simulations of disintegrating satellites can reconstruct the shape,
%appearance, and even metallicity distribution of tidal streams
%\citep{far06}.

Clearly, dwarf galaxies must play some role in building stellar halos
of large galaxies because both the Milky Way (MW) and \object{M31}
exhibit tidal streams \citep[e.g.,][]{iba01a,cho02,gil07} and dwarf
galaxies in various stages of disruption \citep[e.g.,][]{iba94,
  how08}.  However, the chemical abundances of individual stars in
present-day MW dSphs do not match that of the MW halo.  \citet{she01}
found a lower \afe\ for dSph red giant branch (RGB) stars than for MW
halo RGB stars.  The differences in chemical abundances led
\citet{tol03} to conclude that the bulk of the halo can not be
composed of stars like those present in surviving dSphs.

In order to compare the bulk metallicities of stars in dSphs and the
MW halo, \citet[hereafter H06]{hel06} obtained medium-resolution
spectra of the MW dSphs Sculptor, Fornax, Sextans, and Carina.  Using
an empirical relation between the infrared \ion{Ca}{2} triplet (CaT)
equivalent width (EW) and \feh\ \citep{tol01}, H06 find a lack of
$\rm{[Fe/H]} < -3.0$ stars in these four MW dSphs. Given this absence,
they concluded that the MW halo, which contains numerous such stars,
could not have formed from present-day dSphs.  Nevertheless, dSphs
could still be the building blocks if they once contained a population
of extremely metal-poor stars. However, it remains difficult to
understand how they could have lost \emph{all} of those stars by now.

In this Letter, we revisit the absence of metal-poor stars in MW dSphs
by targeting lower luminosity galaxies and by using a more direct
technique to measure \feh\ (based on Fe lines) than has been used
before on low or moderate resolution spectra.  In
\S\,\ref{sec:measure}, we describe new metallicity measurements from
the \citet[hereafter SG07]{sim07} data set of eight of the least
luminous MW dSphs.  As a result, we report for the first time the
discovery of extremely metal-poor stars ($\mathfeh < -3.0$) in MW
dSphs.  In \S\,\ref{sec:mdf}, we compare the ultra-faint dSph MDF to
the halo MDF.  We also present the luminosity-metallicity relation for
the full range of MW dSph luminosities.  In \S\,\ref{sec:concl}, we
briefly summarize our findings and discuss work on dSph chemical
abundances beyond [Fe/H].

%%%%%%%%%%%%%%%%%%%%%%%%%%%%%%%%%
%%%%%%%%%   SECTION 2   %%%%%%%%%
%%%%%%%%%%%%%%%%%%%%%%%%%%%%%%%%%

\section{Metallicity Measurements}
\label{sec:measure}
We make use of the observations of SG07, who targeted eight of the
ultra-faint dSphs discovered with SDSS: Coma Berenices, Canes Venatici
I and II, Hercules, Leo IV, Leo T, and Ursa Major I and II.  In
summary, SG07 used DEIMOS on the Keck~II telescope to obtain spectra
at $R \sim 6000$ over a spectral range of roughly 6500--9000~\AA.
[See \citet{guh06} for details on the spectrograph configuration.]  S/N
varied widely from 5--120~\AA$^{-1}$.

\subsection{Technique}
Many previous abundance studies of large samples ($>10$ RGB stars) of
medium-resolution spectra in MW dSphs \citep[H06,
  SG07,][]{win03,tol04,koc06,koc07a,koc07b,bos07,bat06,bat08} have
relied on empirical relations between the CaT and [Fe/H].  These
linear relations are calibrations based on globular clusters (GCs) in
the metallicity range $-2.1 \la \mathfeh \la -0.6$
\citep[e.g.,][]{rut97} or individual stars in MW dSphs with a minimum
$\mathfeh = -2.5$ \citep{bat08}.  All of these calibrations have been
shown to be accurate in their calibrated metallicity ranges.  However,
all of these relations are defined such that they produce a
metallicity floor at $\mathfeh = -2.5$ to $-3.5$, depending on
absolute magnitude, even for a star with no \ion{Ca}{2} absorption.
Therefore, the linear CaT relations must fail at very low
metallicities.  In fact, 56\% of the stars presented here have
absolute magnitudes that yield minimum $\mathfeh_{\mathrm{CaT}}$ (for
zero CaT EW) above $-3.0$.  We choose to employ a different technique
to avoid the much debated issue of the metallicity at which the CaT
method becomes non-linear.

%Even for a spectrum completely free of \ion{Ca}{2} absorption, these
%relations would give metallicities between $\mathfeh = -3.5$ and
%$-2.6$ depending of the star's location on the RGB.  The metallicity
%below which the CaT begins to overpredict \feh\ has not yet been
%found, although \citet{koc08} find that the CaT overpredicts the
%metallicity of one star at $\mathfeh = -2.50 \pm 0.19$ in the Carina
%dSph by 0.44~dex.

\citet*[hereafter KGS08]{kir08} describe a technique to measure
metallicities from moderate resolution, far-red spectra of RGB stars.
The method compares an observed spectrum to a grid of synthetic
spectra at a range of effective temperatures, surface gravities, and
compositions.  Given a photometric estimate of temperature and
gravity, the [Fe/H] of the synthetic spectrum with the best
pixel-to-pixel match to the observed spectrum is adopted as the
measured [Fe/H] for the star.  This approach is similar to a
high-resolution spectroscopic abundance analysis.

\begin{figure}[t!]
%\plotone{f1.eps}
\includegraphics[width=\columnwidth]{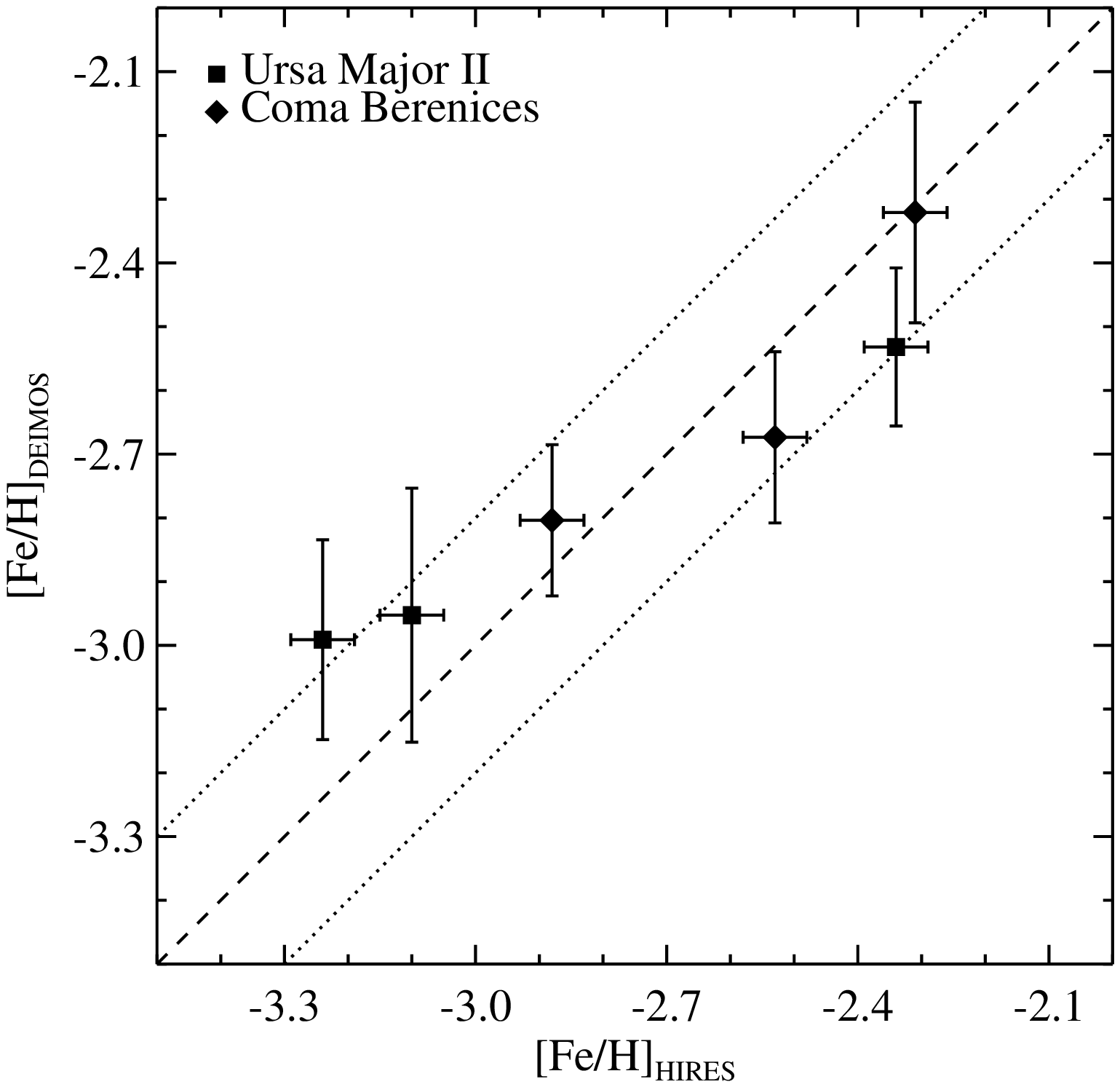}
\caption{Metallicities for the RGB stars in two ultra-faint dSphs
  observed with both HIRES and DEIMOS.  The $x$-axis shows the high
  resolution spectroscopic metallicities.  The $y$-axis shows medium
  resolution synthetic metallicities from this work, which are
  consistent with the HIRES metallicities.  The dashed line is
  one-to-one, and the dotted lines are at $\pm 0.2$~dex to guide the
  eye.
  \label{fig:fehcompare}}
\end{figure}

We exclude blue horizontal branch stars and spectra for which the S/N
was too low to permit a radial velocity cross-correlation measurement
($\mathrm{S/N} \la 10~\mathrm{\AA}^{-1}$).  We transform SDSS $gri$
magnitudes to Johnson-Cousins $VI$ magnitudes following \citet{cho08}
in order to derive the temperatures (\teff) and gravities (\logg) in
the same way as KGS08.  The results are unaffected by using
alternative photometric transformations.  To avoid any systematic
effects of varying [$\alpha$/Fe] ratios on our [Fe/H] measurements, we
mask all the spectral regions susceptible to absorption by Mg, Si, S,
Ar, Ca, or Ti.  These regions comprise 18\% of the spectral range.
Future work will address [$\alpha$/Fe] abundance ratios for these
galaxies.

\subsection{Assessments of the metallicities}
\label{sec:assess}

\begin{figure}[t!]
%\plotone{f2.eps}
\includegraphics[width=\columnwidth]{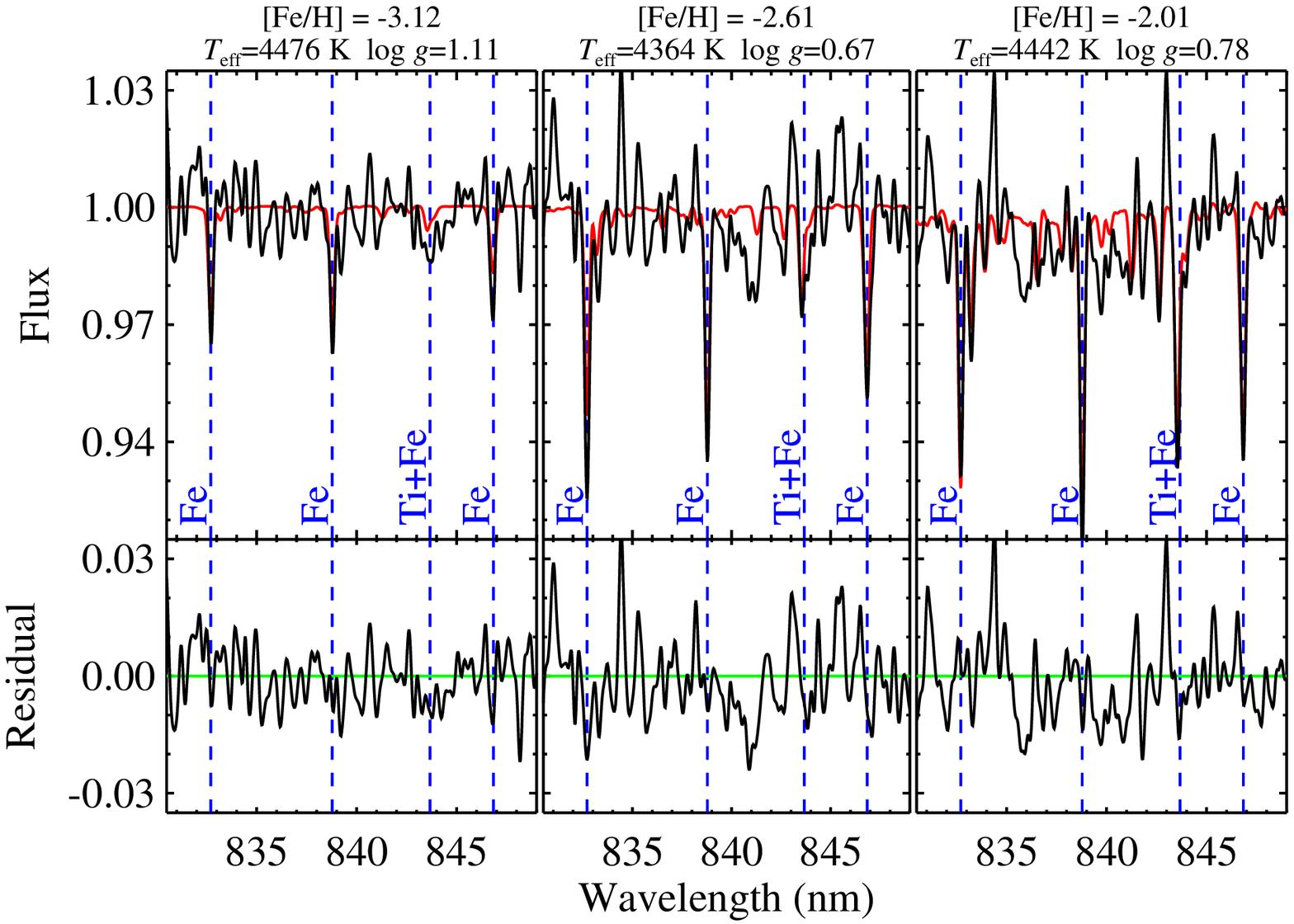}
\caption{Small portions of DEIMOS spectra from three example stars at
  similar effective temperatures ({\it black}).  The spectra are
  smoothed for clarity.  The unsmoothed absorption line depths are
  about 2.3 times greater than shown here.  Lines of Fe~\textsc{i} and
  a Ti+Fe blend are labeled.  The red lines in the top panels are
  synthetic spectra corresponding to the \feh, \teff, and \logg\ of
  the observed stars.  In the bottom panels, the black lines are the
  differences between the observed and synthetic spectra, and the
  green lines are zero.\label{fig:examples}}
\end{figure}

KGS08 demonstrate their technique on Galactic GCs and show that [Fe/H]
may be determined as accurately as 0.1~dex on high S/N spectra and
$\sim$0.5~dex on spectra with S/N as low as 10~\AA$^{-1}$.  The most
metal-poor system that KGS08 analyzed is \object{M15} ($\mathfeh =
-2.4$).  To test this technique at lower metallicities, we compare our
[Fe/H] abundances to those determined from new Keck/HIRES spectra of
several stars in UMaII and Com.  The high-resolution abundance
analysis of these stars (Frebel et al., in prep.)\ shows that the
KGS08 method accurately reproduces these numbers at least down to
$\mathfeh \sim -3.0$.

Figure~\ref{fig:fehcompare} compares synthetic metallicities to HIRES
metallicities and demonstrates excellent agreement.  We emphasize that
the KGS08 technique is a direct measurement of a star's iron and
iron-peak absorption lines and does not make use of any calibration to
obtain metallicities.  Therefore, it is technically not restricted to
any metallicity range.

Figure~\ref{fig:examples} shows three example spectra at three
different metallicities.  The spectra were chosen to have similar
\teff\ and relatively high S/N.  Synthetic spectra are also shown, as
well as the residuals, which scatter evenly about zero except for a
few upward spikes that coincide with incompletely subtracted sky
emission lines.  These examples demonstrate the ability for neutral
iron lines to discriminate easily between stars with different (very
low) metallicities even at moderate spectral resolution.

%%%%%%%%%%%%%%%%%%%%%%%%%%%%%%%%%
%%%%%%%%%   SECTION 3   %%%%%%%%%
%%%%%%%%%%%%%%%%%%%%%%%%%%%%%%%%%

\section{[Fe/H] Distributions of the Ultra-Faint dSphs}
\label{sec:mdf}

\begin{deluxetable}{lrcccc}
\tablecolumns{6}
\tablecaption{Ultra-Faint dSph Metallicities\label{tab:met}}
\tablehead{\colhead{dSph}  &  \colhead{$N$\tablenotemark{a}}  &  \colhead{$\log (L/L_{\sun})$\tablenotemark{b}}  &  \colhead{$\langle\mathrm{[Fe/H]}\rangle$}  &  \colhead{$\sigma_{\mathrm{[Fe/H]}}$}  &  \colhead{$\log (\mathrm{S/N})$\tablenotemark{c}}}
\startdata
UMaII   &  12 & $3.6 \pm 0.2$ & $-2.44 \pm 0.06$ & $0.57$ & $1.5 \pm 0.4$ \\
LeoT    &  19 & $5.1 \pm 0.3$ & $-2.02 \pm 0.05$ & $0.54$ & $1.1 \pm 0.2$ \\
UMaI    &  28 & $4.1 \pm 0.1$ & $-2.29 \pm 0.04$ & $0.54$ & $1.5 \pm 0.3$ \\
LeoIV   &  12 & $3.9 \pm 0.2$ & $-2.58 \pm 0.08$ & $0.75$ & $1.3 \pm 0.3$ \\
Com     &  24 & $3.6 \pm 0.2$ & $-2.53 \pm 0.05$ & $0.45$ & $1.5 \pm 0.3$ \\
CVnII   &  16 & $3.9 \pm 0.2$ & $-2.19 \pm 0.05$ & $0.58$ & $1.5 \pm 0.1$ \\
CVnI    & 165 & $5.4 \pm 0.1$ & $-2.08 \pm 0.02$ & $0.46$ & $1.3 \pm 0.3$ \\
Herc    &  22 & $4.6 \pm 0.1$ & $-2.58 \pm 0.04$ & $0.51$ & $1.6 \pm 0.4$ \\
\enddata
\tablecomments{Data for individual stars (RA, Dec, $V$, $I$,
  $\mathrm{EW}_{\mathrm{CaT}}$, and $\mathfeh_{\mathrm{synth}}$)  are
  available on request from the first author.}
\tablenotetext{a}{Number of member stars, confirmed by radial
  velocity, with measured [Fe/H].  This number is less than the total
  number in SG07 because we exclude spectra with $\mathrm{S/N} \la
  10~\mathrm{\AA}^{-1}$.}
\tablenotetext{b}{We adopt luminosities of \citet{mar08} except for
  LeoT, for which we adopt the luminosity of \citet{dej08}.}
\tablenotetext{c}{Average spectral signal-to-noise ratio per \AA.}
\end{deluxetable}

For each dSph that SG07 observed, Table~\ref{tab:met} shows the number
of stars we analyze, mean \feh, the rms dispersion in \feh, and the
distribution of S/N.  The individual stellar \feh\ values are then
weighted by the inverse square of the errors and averaged to obtain
the mean \feh\ value of every dSph.  These metallicity measurements
establish some of our dSphs as the least enriched known stellar
systems except for the MW halo, and more metal-poor than any GC.

%In comparison to the CaT metallicities of SG07, the mean
%metallicities we measure for the dSphs are $> 0.2$~dex lower for five
%of the dSphs.  Additionally, we find much larger dispersions in
%[Fe/H] than SG07.

\subsection{Metal-poor tail}

We have discovered for the first time stars in dSphs that are more
metal-poor than $\mathfeh = -3.0$.  Our sample contains 15 such stars
in seven dSphs.  Only UMaII contains no stars with
$\mathfeh_{\mathrm{DEIMOS}} < -3.0$, although it does contain two
stars with $\mathfeh_{\mathrm{HIRES}}= -3.1$ and $-3.2$.  To assess
the significance of these detections, we compare the metal-poor tail
of the MDF for all the ultra-faint dSphs to the Fornax MDF of H06.  We
choose a conservative metallicity cut of $\mathfeh < -2.0$, which
includes 178 ultra-faint dSph stars and 83 Fornax stars.  For each
star in the metal-poor tail of the SG07 sample, we randomly select one
counterpart from the metal-poor tail in Fornax. To account for the
different sample sizes, some Fornax stars are used more than once.  We
then randomly resample the H06 $\mathfeh_{\mathrm{CaT}}$ measurement
for each Fornax star from a normal distribution whose standard
deviation is given by the uncertainty ($\delta\mathfeh$) for its
counterpart in the SG07 sample. We find that, at $\mathfeh < -2.0$,
$\delta\mathfeh$ does not vary with metallicity.

After each star in the SG07 sample has been paired in this way, we
count the number of stars with $\mathfeh < -3.0$ in the resampled
Fornax distribution.  With $10^6$ Monte-Carlo resampling realizations,
the number distribution of stars with $\mathfeh < -3.0$ appears
roughly Poissonian with a median frequency of 5.  Just 47 realizations
contained at least 15 stars with $\mathfeh < -3.0$.  Therefore, we
conclude that the probability that our detection of 15 stars with
$\mathfeh < -3.0$ is consistent with being drawn from the H06 Fornax
MDF is very low: $P_{\mathrm{For}} = 4.7 \times 10^{-5}$.  We repeat
this test with the other three H06 MDFs to determine $P_{\mathrm{Scl}}
= 3.4 \times 10^{-5}$, $P_{\mathrm{Car}} = 2.5 \times 10^{-5}$, and
$P_{\mathrm{Sex}} = 4.1 \times 10^{-3}$.  Although Sculptor contains
the lowest single-star \feh\ measurement (\mbox{$\mathfeh = -2.86$})
in the sample of H06, the Sextans MDF is most heavily weighted toward
low metallicities and therefore has the highest probability of
consistency with our findings of $\mathfeh < -3.0$ stars.  These
statistical tests are quite conservative and do not consider that we
actually detect stars as low as $\mathfeh = -3.3$.

%\subsection{Comparison to MW Halo}
%\label{sec:halo}

\begin{figure}[t!]
%\plotone{f3.eps}
\includegraphics[width=\columnwidth]{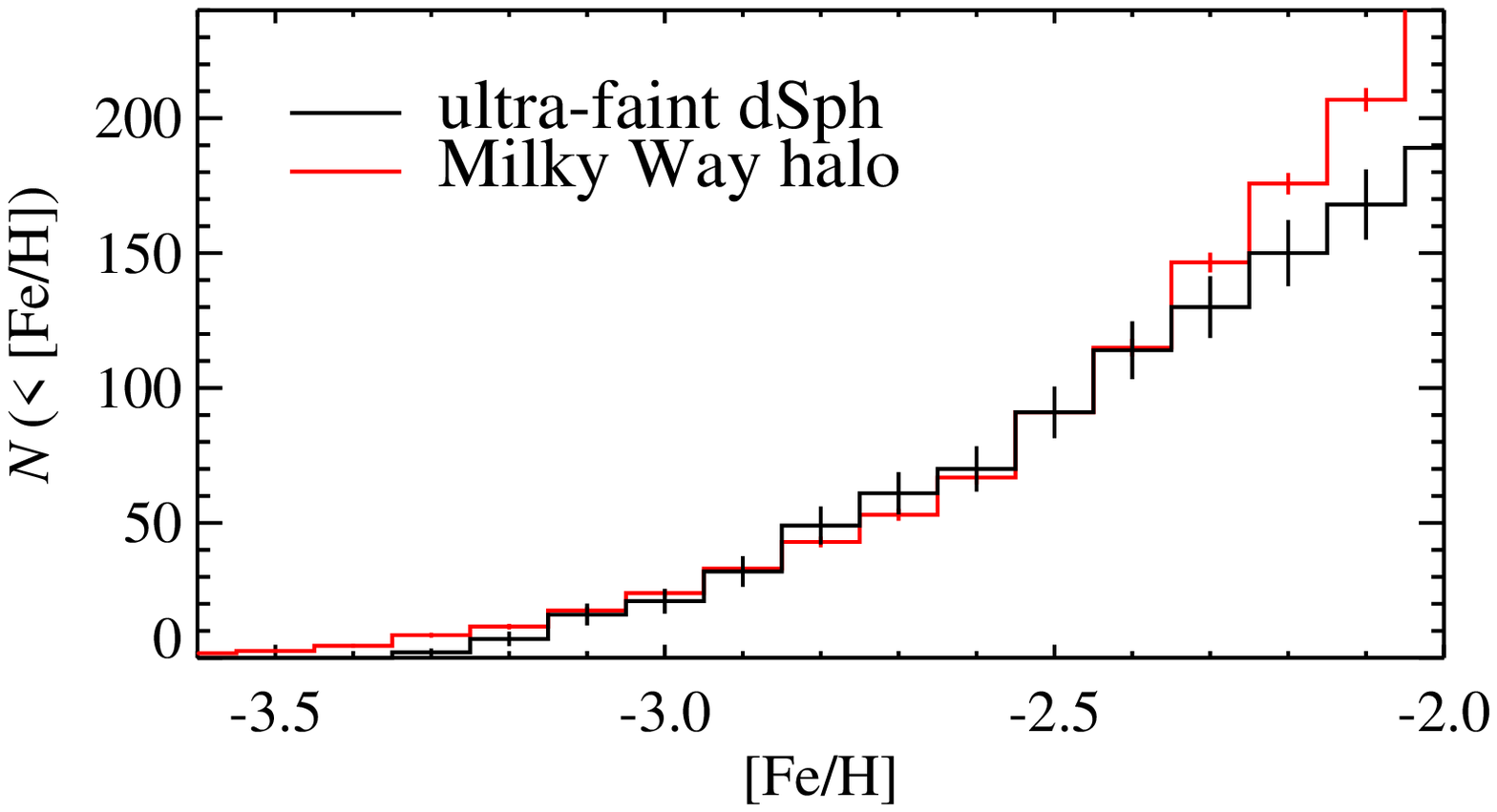}
\caption{Cumulative MDFs for the metal-poor tails of the eight
  ultra-faint dSphs ({\it black}) and the MW halo \citep[{\it
      red},][]{bee05}.  The red histogram is normalized to contain the
  same number of stars with $\mathfeh < -2.45$ as the black histogram.
  The error bars are Poissonian.\label{fig:mdf_mp}}
\end{figure}

In light of the claim by H06 that For, Scl, Car, and Sex lack the
metal-poor tail of stars that is present in the MW halo, we compare
our newly measured MDF of the ultra-faint dSphs to the MW halo MDF.
Figure~\ref{fig:mdf_mp} shows the metal-poor end of the halo
cumulative MDF from the HK and Hamburg/ESO Surveys with
carbon-enhanced stars removed \citep{bee05} compared to the cumulative
MDF for all eight ultra-faint dSphs combined.  The halo histogram is
normalized to the number of ultra-faint dSph stars with $\mathfeh <
-2.45$ in order to mute the incompleteness of the halo MDF at higher
\feh.  This rough comparison can be done more rigorously when a more
complete halo MDF becomes available. In the meantime, we find that the
shape of the metal-poor halo MDF agrees qualitatively with that of the
ultra-faint dSph MDF.  Note that the latter MDF covers a narrower dSph
luminosity range than the broad range of dwarf galaxies which
presumably built the stellar halo.  As a result, the ultra-faint dSph
MDF will cover a narrower metallicity range than the halo MDF because
of the different star formation efficiencies.

\begin{figure}[t!]
%\plotone{f4.eps}
\includegraphics[width=\columnwidth]{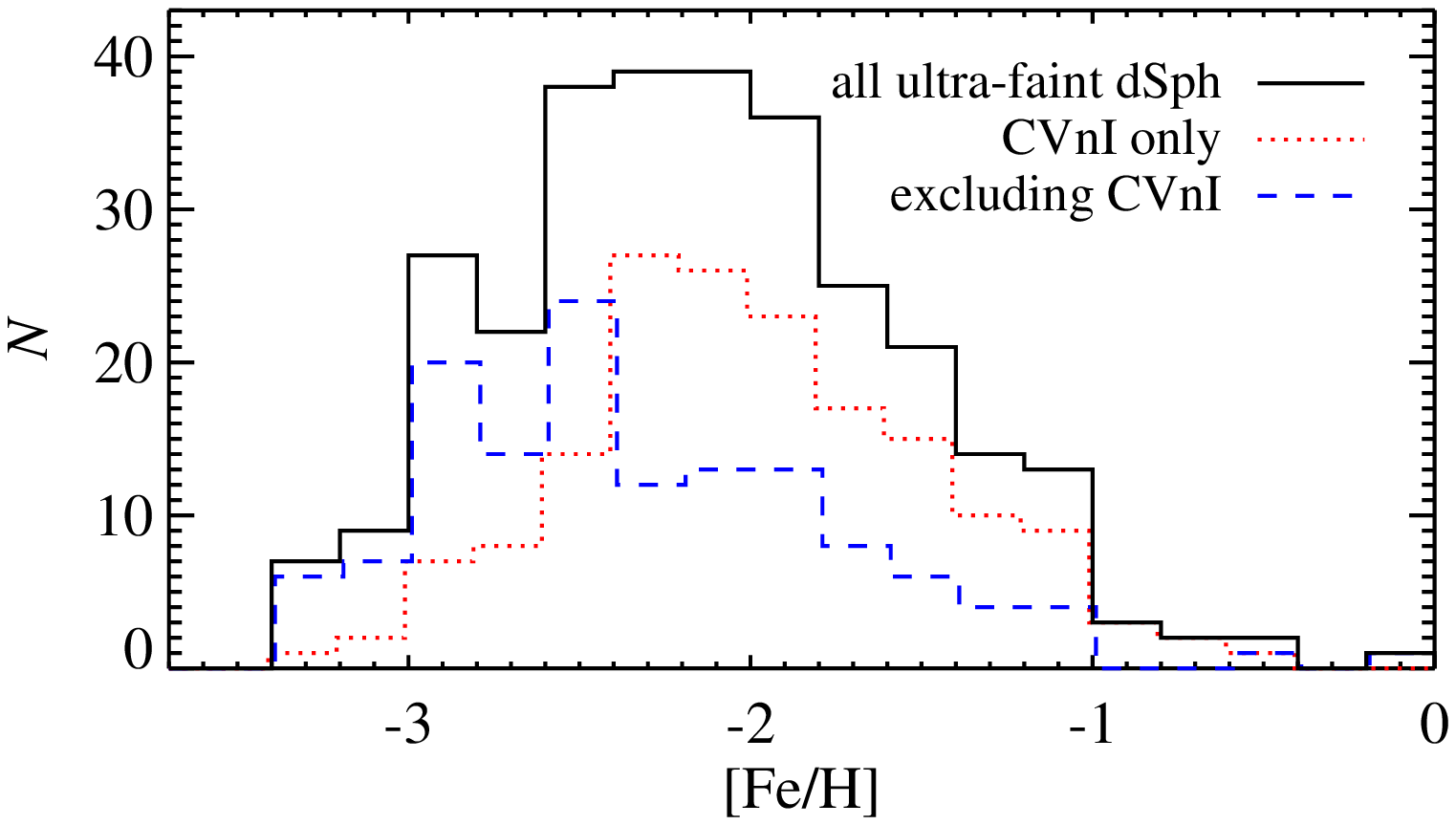}
\caption{Combined MDF for all eight ultra-faint dSphs ({\it black}),
  for CVnI only ({\it red dotted}), and for all ultra-faint dSphs
  except CVnI ({\it blue dashed}).  CVnI is the most luminous
  satellite of those presented here, and it is also the most
  metal-rich.\label{fig:mdf}}
\end{figure}

Figure~\ref{fig:mdf} shows the combined MDF for all eight dSphs,
spanning the range $-3.3 < \mathfeh < -0.1$.  Because CVnI is
significantly more luminous and more metal-rich, we also display its
MDF separately.

\subsection{Luminosity-metallicity relation}
\label{sec:lz}

\begin{figure}[t!]
%\plotone{f5.eps}
\includegraphics[width=\columnwidth]{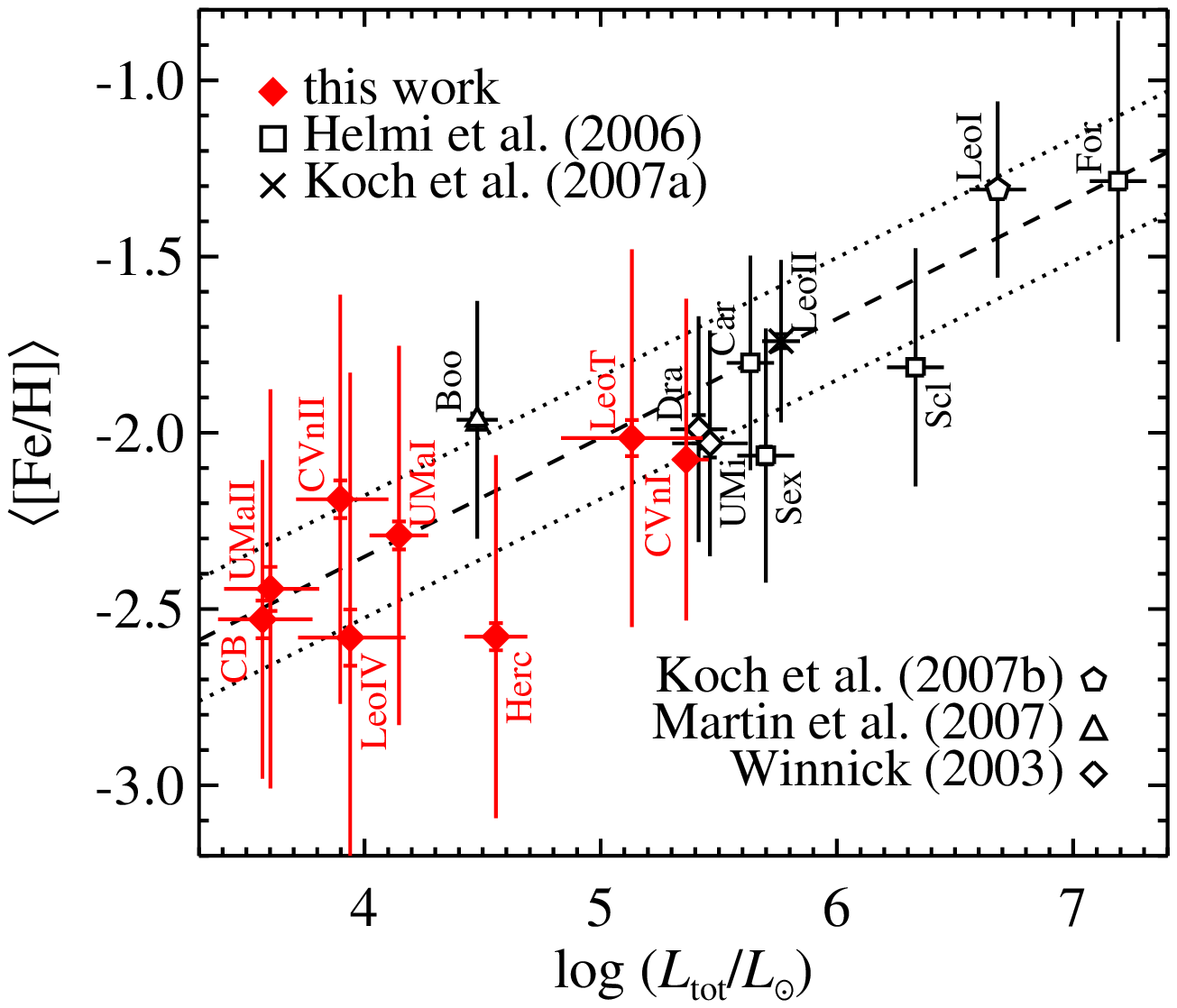}
\caption{The mean [Fe/H] of MW dSphs vs.\ total luminosity.  For those
  dSphs not listed in Table~\ref{tab:met}, we adopt the luminosities
  of \citet[][Boo]{mar08} and \citet[][others]{mat98}.  The figure
  legends give the sources of \feh\ measurements.  The dashed line is
  the weighted, least-squares straight line fit in $\log (L)$-[Fe/H]
  space, accounting for the errors in both $L$ and \feh\
  \citep{akr96}.  The dotted lines are the rms dispersion of the
  residuals.  The full vertical error bars are the rms dispersions of
  \feh\ within a single galaxy, and the hatch marks (not visible for
  all dSphs) are the errors on $\langle\mathrm{[Fe/H]}\rangle$.  The
  MW satellite luminosity-metallicity relation is well-defined for
  nearly 4~dex in luminosity.\label{fig:fehlum}}
\end{figure}

The segregation by luminosity of the ultra-faint dSphs, combined with
our new [Fe/H] measurements, leads us to re-determine the
luminosity-metallicity relation for all MW dSphs except Sagittarius,
which is a very metal-rich outlier. We also exclude the least luminous
objects (Willman 1, Segue 1, and Bo{\" o}tes II) because they have
only a few RGB stars, and their metallicities are not well known.
Figure~\ref{fig:fehlum} combines \feh\ and $L$ for the classical dSphs
with data for the ultra-faint dSphs (Table~\ref{tab:met}).  Over the
full 3.6~dex range of luminosity, this combined sample shows a
well-defined relation.  Our ultra-faint dSphs extend the trend found
in the more luminous systems.

The following equation describes the fit, where the errors are the
standard deviations of the slope and intercept:

\begin{displaymath}
\langle\mathrm{[Fe/H]}\rangle = (-2.01 \pm 0.05) + (0.34 \pm 0.05)
\log \left(\frac{L_{\mathrm{tot}}}{10^5 L_{\sun}}\right)
\end{displaymath}

\noindent
The linear Pearson correlation coefficient for the data is 0.89,
indicating a highly significant correlation.

%%%%%%%%%%%%%%%%%%%%%%%%%%%%%%%%%
%%%%%%%%%   SECTION 4   %%%%%%%%%
%%%%%%%%%%%%%%%%%%%%%%%%%%%%%%%%%

\section{Conclusions}
\label{sec:concl}

We have presented metallicity measurements for eight of the least
luminous known galaxies in the Universe.  We also discover, for the
first time, stars outside the MW field halo population with $\mathfeh
< -3.0$.  Furthermore, we have shown that the distribution of [Fe/H]
in present-day dSphs reaches nearly as low as that of the MW stellar
halo, and that the dSph luminosity-metallicity relation is
well-defined for nearly 4~dex in luminosity.

There are two main differences between our study and previous works
that might have contributed to our discovery of extremely metal-poor
stars in MW dSphs.  First, our spectral synthesis approach is valid
for any metallicity and is not restricted to calibrated ranges, like
the CaT technique.  Second, we explored very faint dSphs whereas H06
examined more luminous dSphs. We have found extremely metal-poor stars
only in the faintest dSphs. It remains to be seen whether any of the
brighter dSphs also harbor extremely metal-poor stars.

[Fe/H] is just one abundance puzzle in the role of dSphs in building
the stellar halo.  Additional elements will need to be examined to
obtain further clues. Most notably, [$\alpha$/Fe] ratios in the more
luminous dSphs are on average lower than in the halo
\citep[e.g.,][]{she03,gei07}.  However, cosmologically motivated
models including star formation and chemical enrichment
\citep{rob05,fon06} may explain the difference.  These models along
with the discovery of extremely metal-poor stars in long-lived dSphs
support the original hierarchical paradigm of galaxy formation
\citep[e.g.,][]{sea78,whi78}.

\acknowledgments
We thank Jennifer Johnson, Kim Venn, and the referee for valuable
advice.  We acknowledge National Science Foundation grants AST-0307966
and AST-0607852 and NASA/STScI grants GO-10265.02 and GO-10134.02.
E.~N.~K.\ is supported by a NSF Graduate Research Fellowship.
A.~F.\ acknowledges support through the W.~J. McDonald Fellowship of
the McDonald Observatory.

{\it Facility:} \facility{Keck:II (DEIMOS)}

\end{document}